\documentclass[fleqn,twoside]{article}
\usepackage{espcrc2}

\usepackage{graphicx}
\usepackage{epsfig}
\usepackage[figuresright]{rotating}

\newcommand{\AmS}{{\protect\the\textfont2
  A\kern-.1667em\lower.5ex\hbox{M}\kern-.125emS}}

\title{Measurement of the Muon Anomalous Magnetic Moment to 
0.7 ppm~\cite{proc}}

\author{Yannis K. Semertzidis\address[BNL]{Brookhaven National Laboratory, 
Upton, New York 11973   \\ 
        for the g-2 collaboration~\cite{bennett02}}%
        }
       
\begin{document}

\begin{abstract}
The experimental method  together with the analysis method and 
results of the data taken in 2000  and prospects
 of the muon anomalous magnetic and electric dipole moment experiments are
presented here. 
\vspace{1pc}
\end{abstract}

\maketitle

\section{Introduction}

The $g$ factor of a particle, defined as the ratio of the magnetic moment of the
particle in units of its Bohr magneton, over the angular momentum of the
same particle in units of $\hbar$:

\begin{equation}
g = {{\rm magnetic \,\, moment \over e \hbar/2 m c} \over 
{\rm angular \,\, moment \over \hbar}}, \label{eq:g-factor}
\end{equation}

 \noindent is used  to enhance our understanding of the underlying
theory, like:

\begin{enumerate}

\item{}  It is used to indicate that the proton ($g_p = +5.586$) and the neutron 
($g_n = -3.826$) are composite particles.

\item{} The ratio $g_p / g_n$ of -1.46 being close to the predicted -3/2
 was the first success of the constituent quark model.

\item{} The g-2 value of the electron is non-zero due to quantum field 
fluctuations.  The agreement between the experimental  and the theoretical
value is a triumph of both QED and of the experimental approach.

\item{} The g-2 value of the muon is usually 
more  sensitive to higher mass particles
than the electron g-2 by the ratio of $(m_\mu /m_e)^2 \approx 40000$.
Therefore it is used widely to check the validity of the standard model
and as a sensitive probe for physics beyond it.

\end{enumerate}

\section{Theory}

The anomalous magnetic moment of the muon, defined as $a_\mu = {g_\mu - 2 \over 2}$,
is the sum of QED, hadronic, and weak interaction contributions plus any
new physics that may be present:

\begin{eqnarray}
a_\mu({\rm theo}) & = & a_\mu({\rm QED}) + a_\mu({\rm had}) + a_\mu({\rm weak}) + \nonumber \\
              &   & a_\mu({\rm new \, \, physics}).
\end{eqnarray}

The theoretical values of the various contributions, especially that
of the hadronic one,  are the subject of many 
papers~\cite{davier1,hisano,bern,bennett,rafael,teubner,davier2,marciano}:

\begin{itemize}

\item{} $a_\mu({\rm QED}) = 11 \, 658 \, 470.57(0.29) \times 10^{-10} \,\, (0.025 \, \rm ppm)$ ~\cite{mohr}.

\item{} $a_\mu({\rm had}) = 683.3(7.7) \times 10^{-10} \,\, (0.66 \, \rm ppm)$,
based on the $e^+ e^-$ value for the lowest-order hadronic correction
of $a_\mu({\rm had,1}) = 684.7(7.0) \times 10^{-10} \,\, (0.60 \, \rm ppm)$~\cite{davier2}
since only
these data can be directly related to  $a_\mu({\rm had})$ without further
theoretical assumptions.
The higher-order hadronic contribution of
$a_\mu({\rm had,2}) = -10.0(0.6) \times 10^{-10}$~\cite{krause}, and 
$a_\mu({\rm had,lbl}) = 8.6(3.2) \times 10^{-10}$~\cite{knecht}. 
 For completeness $a_\mu({\rm had,1}) = 701.9(6.2)\times 10^{-10} \,\,(0.53 \, \rm ppm)$
 based on the $\tau$ data~\cite{davier2}.

\item{} $a_\mu({\rm weak}) = 15.1(0.4) \times 10^{-10} \,\, (0.03 \, \rm ppm)$~\cite{marciano}.

\end{itemize}

\begin{figure}[p]
\includegraphics*[width=19pc]{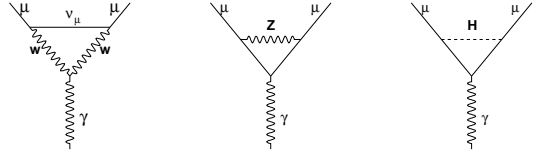}
\vskip -0.8 cm
\caption[weak2ndorder]{{ Second order weak contributions to $a_\mu$ involving the $W$, $Z$, and 
Higgs.  The Higgs contribution is negligible for the present mass limits.  The total 
relative weak contribution is estimated to be +1.3~ppm with an error of 0.03~ppm.}
\label{fg:weak2ndorder}} 
\end{figure}

\begin{figure}[p]
\includegraphics*[width=17pc]{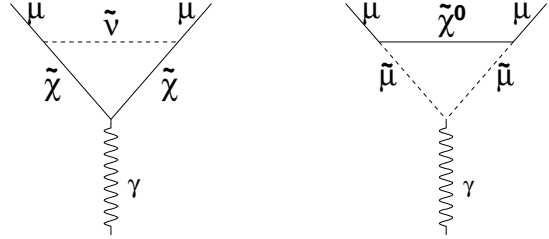}
\vskip -0.8 cm

\caption[beyondsm]{{ SUSY contributions to $a_\mu$ involving the superpartners of $W$, and 
$Z$.  Their contribution, enhanced by the factor $\tan{\beta}$, may be larger than the weak.}
\label{fg:beyondsm}} 
\end{figure}

Then  the standard model contribution
$ a_\mu({\rm SM}) = 11 \, 659 \, 169(7.7) \times 10^{-10}$ with a relative
uncertainty of $ (0.66 \, \rm ppm)$. 
The Feynman diagrams of the second order weak contributions are shown in Figure~(\ref{fg:weak2ndorder}).
 An example of $a_\mu({\rm new \, \, physics})$ 
is the 
contribution due to SUSY, Figure~(\ref{fg:beyondsm}), where the supersymmetric
partners of $W$ and $Z$, the chargino and neutralino, 
are involved.  Their contribution is 
estimated to be 

\begin{equation}
a_\mu({\rm SUSY}) = 14 \times 10^{-10} \left( {100 \rm GeV \over m_{\rm susy}} \right) ^2 \, \tan{\beta}.
\end{equation}

\section{Experimental Method}

\begin{figure}[p]
\includegraphics*[width=12pc,angle=270]{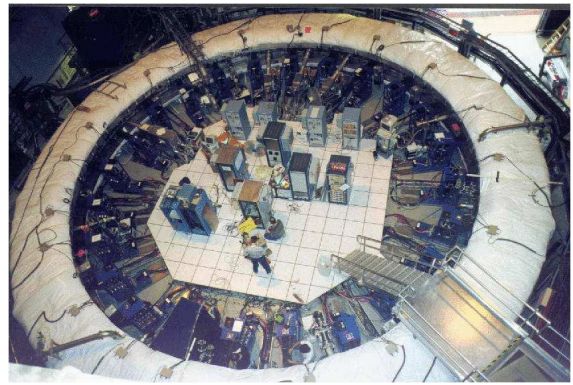}
\vskip -0.7 cm
\caption[field]{{ The g-2 muon storage ring is shown here with the thermal isolation cover
(white) in place.}
\label{fg:ring}} 
\end{figure}

A bunch  of highly polarized muons with momentum $P \approx 3.09 \, {\rm GeV/}c$  
is injected into a ring of 7.112~m 
radius,  
Figure~(\ref{fg:ring}),  with
 1.45~T dipole magnetic field of very high uniformity.  
The time distribution of the injected beam has an r.m.s. of 25~ns.
At $\approx 90^\circ$ 
from the injection point, the muon beam is kicked onto stable orbits by a
fast magnetic pulse (kicker)~\cite{stratos}.
Within a super-cycle of 3.2~s
there are  12 bunches, 33~ms apart from each other. Vertical
focusing is provided by electrostatic quadrupoles~\cite{yannis}.
  Muons of that momentum have a lifetime
of $\gamma \tau_\mu \approx 64.4 \, \rm \mu s$ and are stored in the ring
for $\approx 1.4$~ms.

\subsection{Principle of the g-2 Experiment}
The muon  momentum vector 
in the lab frame, $\vec{P}$, precesses under the influence of the electromagnetic
forces.
The muon spin in its own rest frame, $\vec{S}$,
 precesses  under the influence of only the magnetic
forces present in its  rest frame.  Muon decay violates parity in a maximal way; 
in the muon rest frame the
most energetic electrons go along the muon spin direction.
Then the energy of the electron is Lorentz boosted due to the muon momentum 
resulting in an energy which is modulated according to the dot product of the 
two vectors~\cite{picasso,jackson}, 
$\vec{S} \cdot \vec{P}$, which for a spin 1/2 particle is an exact sine wave
of angular frequency $\omega_a$, giving rise to
the principle of g-2 frequency detection.

\begin{figure}[h]
\includegraphics*[width=17pc]{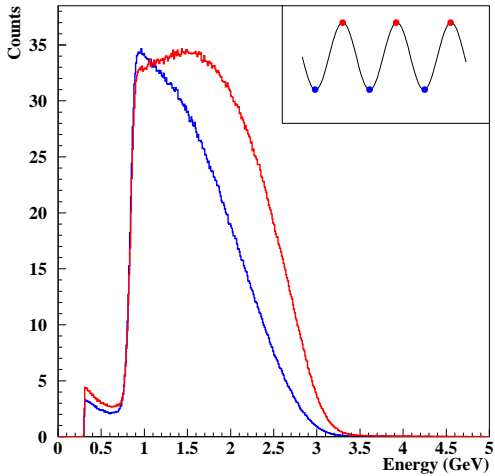}
\vskip -1.0 cm
\caption[energy]{{ The energy spectrum of the detected positrons in red (blue)
 for a muon spin parallel
(anti-parallel) to its momentum in the lab frame. The red line corresponds to
 the  higher energy
positrons.  The inset shows the number of positrons as a function of time for an energy
threshold of 2~GeV.  The red dots correspond to the highest number of detected positrons 
above 2~GeV, whereas the blue to the lowest.}
\label{fg:energy}} 
\end{figure}

The electrons, having on average less momentum than the stored muons,
spiral inwards where they are collected by an inner ring of 24 electromagnetic
calorimeters~\cite{sedykh}, and their signals are recorded by  400~MHz 
waveform digitizers (WFD).   The acceptance of the detectors depends
on the position of the muon beam and is estimated to be about 1/3 over all decays.
In Figure~(\ref{fg:energy}) we show the energy spectrum of the detected positrons
when the muon spin in its rest frame 
is parallel to the muon momentum in the lab frame (the
higher energy spectrum shown in red) and when the muon spin is anti-parallel (the lower
energy spectrum shown in blue).  Counting the number of detected positrons above
 an energy threshold of 2~GeV, one then gets a sine wave with the g-2 frequency
shown in the inset.

The angular frequency $\omega_a$ in the presence of both $E$ and $B$ fields is 
found~\cite{picasso,jackson} to be

\begin{equation}
\vec{\omega}_a= -{e \over m_\mu} \left[ a_\mu \vec{B} + \left( {1 \over \gamma^2 -1} - a_\mu \right) \vec{\beta} \times \vec{E} \right] , \label{eq:omega}
\end{equation}

\noindent where $\vec{\beta} \cdot \vec{E} = \vec{\beta} \cdot \vec{B}=0$ is assumed.  The
electric field influences $\omega_a$ because it advances the muon spin in its own
rest frame (owing to Lorentz transformations, the electric field is partially transformed 
to a magnetic field) and the muon momentum in the lab frame. 
In the case of a realistic detector with finite energy and time resolution, there is 
some level of overlapping pulses (pileup), which produces a second harmonic of $\omega_a$
but also a first harmonic of $\omega_a$.  More about the time development of these 
components and how we deal with 
them in our data is given later in the analysis section.

\subsection{Magic Momentum}
The muon anomalous magnetic moment has a value of approximately 
${\alpha \over 2 \pi} \approx {1 \over 800}$, which is the second-order (dominant) QED
contribution.  Therefore, for $\gamma \approx 29.3$, 
the above equation~(\ref{eq:omega}) reduces to

\begin{equation}
{\omega}_a= a_\mu {e \over m_\mu}  {B}. \label{eq:omega2}
\end{equation}

\begin{figure}[p]
\includegraphics*[width=17pc]{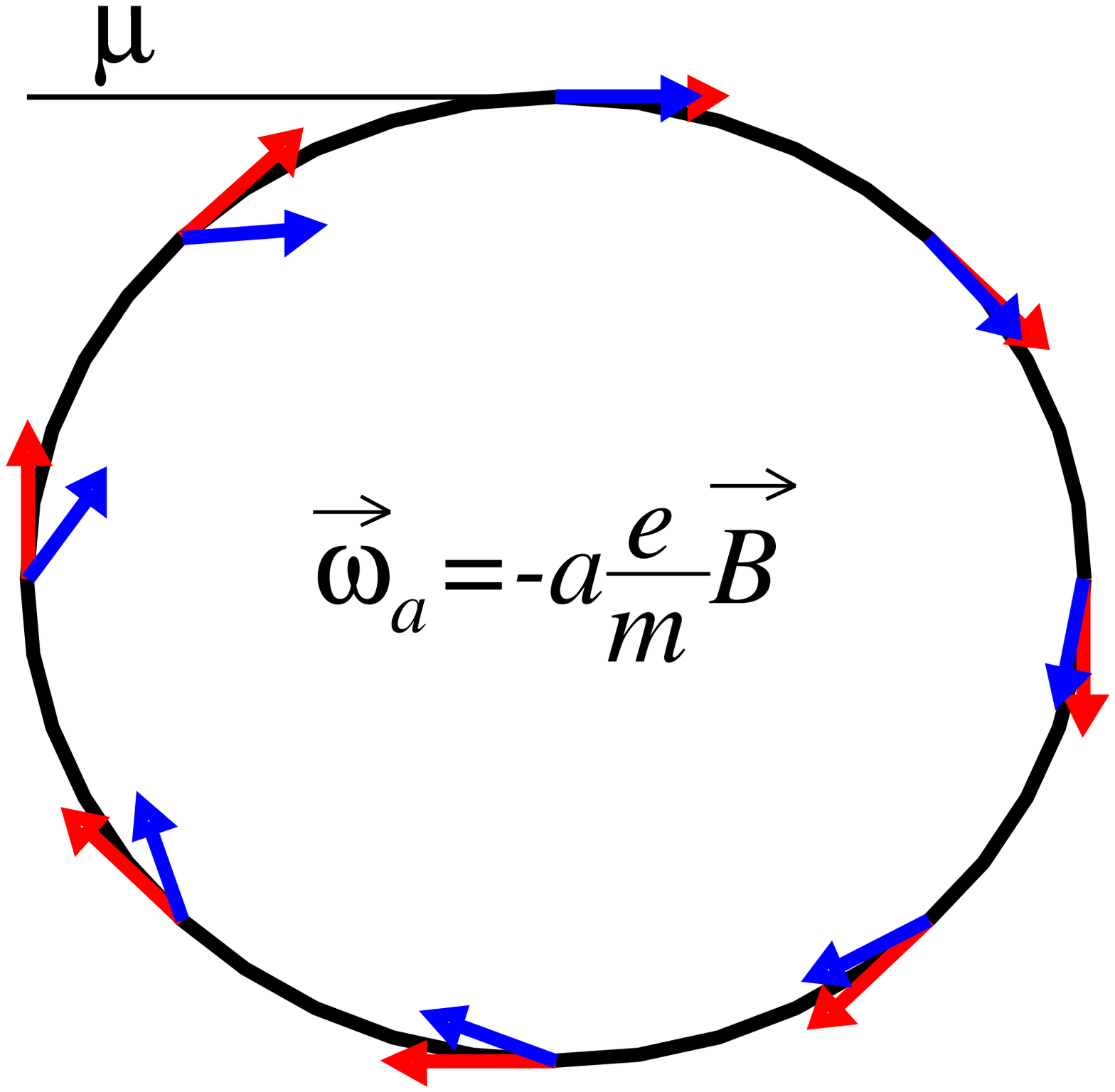}
\vskip -1.0 cm
\caption[field]{{ The muon spin $\vec{S}$ 
in its own rest frame gets ahead of its momentum 
vector $\vec{P}$ in the lab frame as time progresses.  The detected energy spectrum of the
electrons/positrons in the lab frame probes the product $\vec{S} \cdot \vec{P}$.}
\label{fg:spin}} 
\end{figure}

\noindent  In Figure~(\ref{fg:spin}) we show the spin vector getting ahead of the 
momentum vector as a function of time.

The reason equation~(\ref{eq:omega2}) is valid
 comes from the fact that the g-2 precession is the difference between the muon spin 
precession in its own rest frame minus the momentum precession in the lab frame.  The
value of $\gamma \approx 29.3$ corresponds to the case where the radial $E$ field 
precesses the muon spin in its rest frame and momentum in the lab frame 
at the same rate.  This is the reason for choosing
a muon momentum of $P \approx 3.09 \, {\rm GeV/}c$, a.k.a. ``magic momentum''~\cite{cern}, 
which for $B = 1.45$~T corresponds to
a radius of $\approx 7.11$~m.  

\subsection{Origin of the Electric Field and Pitch Corrections}
Due to finite muon beam 
momentum width the cancellation is not
exact and there is a need for a small electric field correction.  Also,
the muon momentum may not be exactly orthogonal to the external
magnetic field, so the Lorentz transformation of the lab electromagnetic fields into
the rest frame fields are slightly modified.  This effect 
introduces a small ``pitch'' 
correction.  Both the $E$ field and pitch corrections are small, their
sum being about +0.8~ppm.

\subsection{Equation to Estimate $a_\mu$}
A muon at rest and while in the presence 
of a magnetic field has its spin precessing with an angular frequency given by

\begin{equation}
\omega_s = {g_\mu \over 2} \left( {e \over m_\mu} \right) B. \label{eq:omega3}
\end{equation}

\noindent When combined with equation~(\ref{eq:omega2}), this yields

\begin{equation}
a_\mu = {\omega_a \over \omega_s - \omega_a } = {R \over \lambda - R}. \label{eq:omega4}
\end{equation}

\noindent Here $R= {\omega_a \over \omega_p}$, and $\omega_p$ is  the angular 
frequency of the free proton in the $B$ field, measured with NMR 
techniques~\cite{ralf,fei}. 
 One then also needs to know 
the ratio $\lambda = {\omega_s \over \omega_p}$, 
the value of which is taken from another
experiment.  It follows that  $\lambda = {\mu_\mu \over \mu_p}$ with $\mu_\mu$  and $\mu_p$
the magnetic moments of muon and free proton respectively.

The precision with which $a_\mu$ can be evaluated is determined by the accuracy of 
 $R$ and $\lambda$. 
 The value of $\lambda$ is determined by measuring the microwave spectrum of 
the ground state of muonium~\cite{liu}, finding
$\lambda=3.183\, 345 \, 39 (10)$.  The precision of $R$ depends on the precision of
$\omega_a$ and $\omega_p$.  The quantity $\omega_a$ is estimated by detecting
the electrons produced by the decay of the muons.  The statistical uncertainty of 
$\omega_a$ is

\begin{equation}
{\delta \omega_a \over \omega_a} = {\sqrt{2} \over \omega_a \gamma \tau_\mu A \sqrt{N_e}}. \label{eq:error}
\end{equation}

\noindent Here $A$ is the g-2 oscillation asymmetry and $N_e$ is the total number of detected
positrons.

\section{Beam Dynamics}

\subsection{Momentum Acceptance of the Muon g-2 Storage Ring}
\begin{figure}[p]
\includegraphics*[width=17pc]{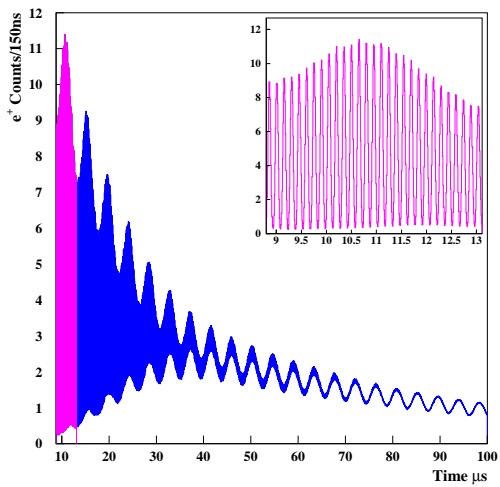}
\vskip -1.0 cm
\caption[fast_orig]{{ The bunched structure of the muon beam, clearly evident in
the inset showing the $ 8.5 - 13 \, \rm \mu s$ time 
range, distorts the g-2 oscillation
at early times. 
The fast structure is decaying with a lifetime of $\approx 25 \, \rm \mu s$ with
only the g-2 oscillation remaining at later times.}
\label{fg:fast_orig}} 
\end{figure}

\begin{figure}[p]
\includegraphics*[width=17pc]{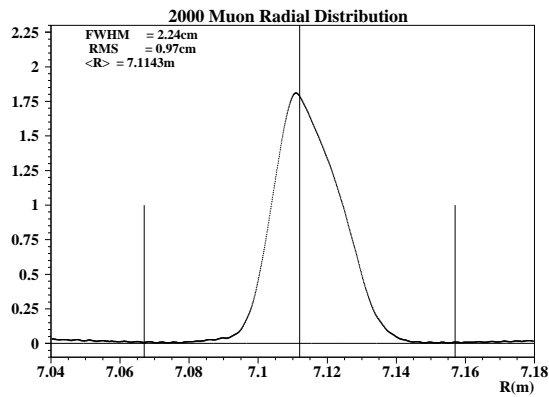}
\vskip -0.7 cm
\caption[momentum]{{ The muon 
radial distribution obtained by Fourier analyzing the fast rotation structure of 
Figure~(\ref{fg:fast_orig}). The central vertical line corresponds to the center of the
muon storage region with $R_0=7.112$~m and the two lines on either side are the edges 
of the storage region at $R_0 \pm 4.5$~cm.}
\label{fg:momentum}} 
\end{figure}

The injected muon beam debunches due to momentum dispersion with a lifetime of 
$\approx 25 \, \rm \mu s$.  The time spectrum of the positrons detected
by a single detector at early times after injection is shown in 
Figure~(\ref{fg:fast_orig}).  The g-2 oscillation along with the slowly decaying
fast rotation structure is clearly shown.  The rotation frequency of the stored muons 
depends on their momentum.  Therefore their momentum distribution can be found by
Fourier analysis of the arrival times of the detected positrons.  The
momentum acceptance of the ring is narrow ($0.6 \%$ total) 
and a special Fourier analysis
technique is required to avoid introducing artificial effects into
the width of the Fourier analyzed data~\cite{yuri}.  The muon radial distribution
so obtained is shown in Figure~(\ref{fg:momentum}).
The muon momentum distribution is inferred from the 
radial distribution using the formula $P_\mu = P_0 \times [1 + (R-R_0)(1-n)/R_0]$,
where $P_0$ is the central momentum, $R_0=7.112$~m is the center of the muon storage
region, $R$ is the actual radial location of the muons, and $n$ is the field focusing index.

\begin{figure}[p]
\includegraphics*[width=17pc]{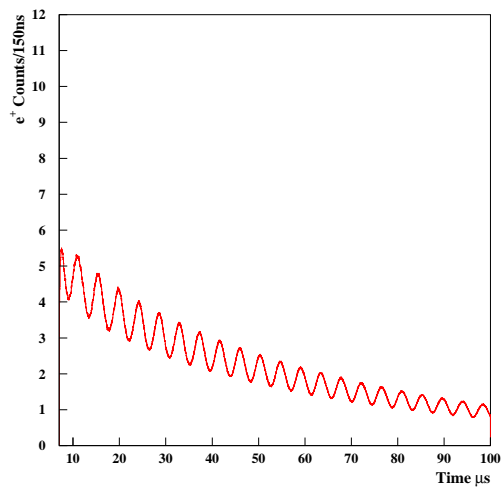}
\vskip -0.8 cm
\caption[fast_random]{{ The detected positron time spectrum of 
Figure~(\ref{fg:fast_orig}) after randomizing
the muon injection time with a period of 149.2~ns; the fast rotation structure 
seen before, is eliminated. }
\label{fg:fast_random}} 
\end{figure}

The fast-rotation structure is eliminated by randomizing the time of the injected
muon beam with the average fast rotation period of approximately 149.2~ns, the result
shown in Figure~(\ref{fg:fast_random}).  After the randomization the Fourier analysis
spectrum shows no structure at the fast rotation frequency.

\subsection{Coherent Betatron Oscillation Frequencies}
\begin{figure}[p]
\includegraphics*[width=17pc]{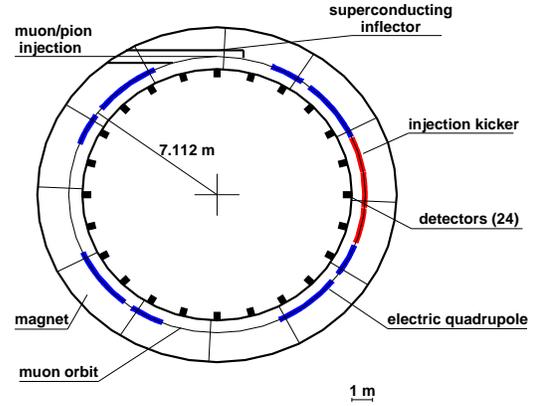}
\vskip -0.8 cm
\caption[lattice]{{The muon g-2 ring  lattice indicating
  the four quadrupole regions, the kicker region and the inflector location.
The muon storage region is a torus of radius  $R_0=7112$~mm and a
cross section of a circle with a diameter of 90~mm. The inner ring of
24  detectors is
shown as black squares.}
\label{fg:lattice}} 
\end{figure}

The muon storage ring lattice is shown in Figure~(\ref{fg:lattice}).  The muon beam
is injected through the inflector~\cite{meng} whose acceptance is smaller than that
of the 
ring itself.  This fact has as a result that the phase space of the betatron 
oscillations 
is not filled, resulting in betatron oscillations of the beam as a whole, called
coherent betatron oscillations (CBO).  Those oscillations are both horizontal and
vertical and their amplitudes are described by:

\begin{equation}
x=x_e + x_0 \cos{(\omega_x t + \theta_x)}, \label{eq:horizontal}
\end{equation}

\begin{equation}
y=y_0 \cos{(\omega_y t + \theta_y)}, \label{eq:vertical}
\end{equation}

\noindent where $x_e$ is the horizontal equilibrium radial position away from the
$R_0=7.112$~m center of the ring, and $x_0$ ($y_0$) is the horizontal (vertical)
 CBO amplitude.  The $\omega_x \equiv 2 \pi f_x$, $\omega_y \equiv 2 \pi f_y$, 
with $f_x$ ($f_y$) the horizontal (vertical) CBO frequencies given by

\begin{equation}
f_x \simeq f_c (1-\sqrt{1-n}), \label{eq:cbo_x}
\end{equation}

\noindent and

\begin{equation}
f_y \simeq f_c \sqrt{n}, \label{eq:cbo_y}
\end{equation}

\noindent where $f_c = {1 \over 149.2 {\rm ns}} = 6.7$~MHz, the cyclotron frequency,
 and
 $n$ is the field focusing index.  Equations~(\ref{eq:cbo_x},\ref{eq:cbo_y}) are not exact
due to the discrete nature of the quadrupole coverage of the ring.
There are small corrections given in reference~\cite{yannis}.

\subsection{Field Focusing Index for the 2000 Run}
\begin{figure}[p]
\includegraphics*[width=18pc]{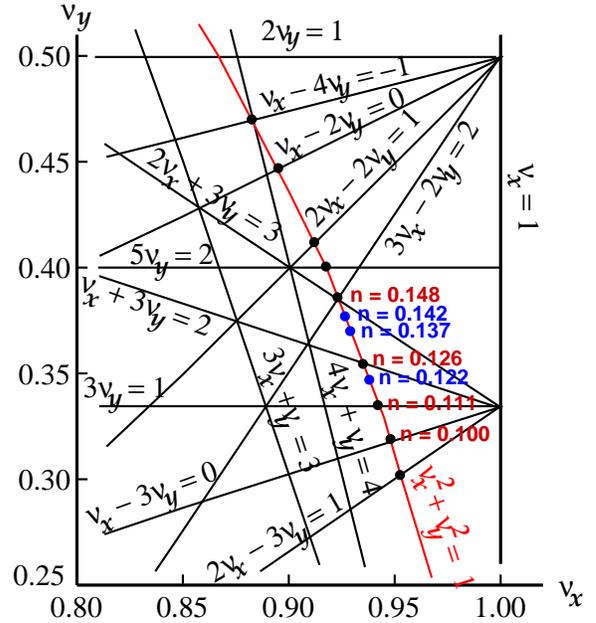}
\vskip -1.0 cm
\caption[tunepl]{{ The tune plane with the most important beam dynamics resonances of our 
weak focusing muon storage ring.  The 2000 run was mostly taken at $n \simeq 0.137$, 
in the middle of
two relatively strong resonances, at
 $n=0.126$ and $n=0.148$.  The 2001 run was taken at 
$n=0.142$ and $n=0.122$.}
\label{fg:tunepl}} 
\end{figure}

In Figure~(\ref{fg:tunepl}) we show the vertical $\nu_y \simeq \sqrt{n}$ 
versus the horizontal $\nu_x \simeq \sqrt{1-n}$ tune, along with
the most important beam dynamics resonances of our weak focusing muon storage ring.  The
acceptance of a weak focusing ring has a rather wide maximum at $n=0.5$, being at
 $90\%$ of its maximum at $n=0.137$.

\begin{figure}[p]
\includegraphics*[width=17pc]{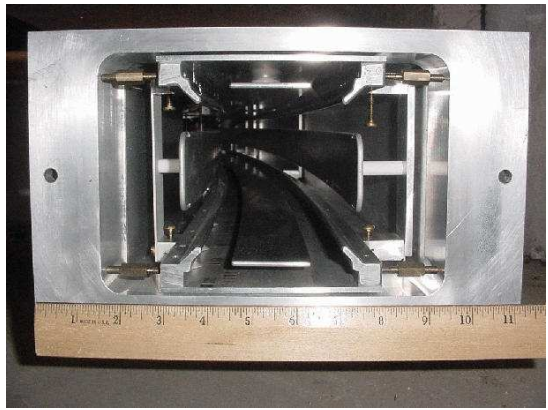}
\vskip -0.5 cm
\caption[quads]{{A photograph of the electrostatic quadrupoles
taken from the end of a vacuum chamber; the ring center is on the left. 
The ruler units  are in inches. }
\label{fg:quads}} 
\end{figure}

The $n$ value is proportional to the voltage  applied to the quadrupole plates,
Figure~(\ref{fg:quads}).  Due to the presence of the
magnetic field, and for reasonable residual pressures in the vacuum chamber of about
$10^{-7}$~Torr,
it would be very difficult to work at $n=0.5$.  This is so because there is a large
number of low-energy trapped electrons circulating in the quad region~\cite{yannis}.
A reasonable number to work with was around $n=0.136-0.137$, 
in the middle between two relatively
strong beam dynamics resonances at $n=0.126$, and $n=0.148$, shown in Figure~(\ref{fg:tunepl}).
The effect of the low-energy trapped electrons has been studied and is shown to 
contribute less than 0.01~ppm  to the magnetic field.  Their effect on the 
quadrupole electric field is negligible~\cite{yannis}.

From equation~(\ref{eq:cbo_x}) we have
 that the horizontal
frequency corresponds to $f_x \simeq 466$~kHz
which is very near twice the g-2 frequency of $\approx 229.1$~kHz.

\section{Analysis of $\omega_a$} 
\begin{figure}[p]
\includegraphics*[width=18.pc]{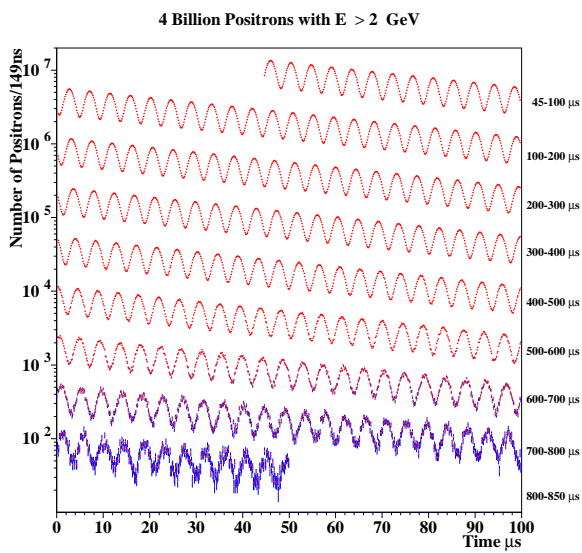}
\vskip -0.8 cm
\caption[wiggles]{{ The number of positrons detected with $E > 2$~GeV as a 
function of time.  The error bars are shown in blue and the points in red.}
\label{fg:wiggles}} 
\end{figure}

In the year 2000, we had a very successful run in terms of accumulating a lot of statistics.
In Figure~(\ref{fg:wiggles}) we show the total number of positrons detected  
with $E > 2$~GeV as a function of time.  The equation describing the ideal positron
time spectrum is given by

\begin{equation}
N(t) = N_0(E) \, e^{- \gamma \tau_\mu} \, [1 + A(E) \, \cos{(\omega_a t + \phi_a(E))}], \label{eq:decay}
\end{equation}

\noindent where $A(E)$ corresponds to the g-2 oscillation asymmetry and $\phi_a(E)$ the 
g-2 phase, both of which depend on the energy threshold $E$; for $E=2$~GeV, $A=0.4$.

\begin{figure}[p]
\includegraphics*[width=17pc]{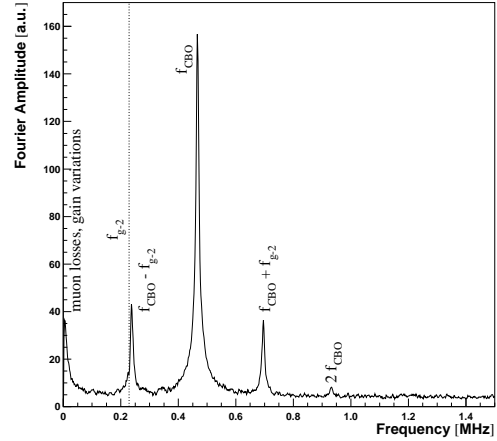}
\vskip -0.8 cm
\caption[fourier]{{ Fourier analysis of the residuals to the fits to the data of 
Figure~(\ref{fg:fourier}) using equation~(\ref{eq:decay}).  $f_{\rm CBO} \equiv f_x$ has
a large amplitude implying that the CBO modulation is important.}
\label{fg:fourier}} 
\end{figure}

The Fourier analysis of the residuals of the fits to the data using equation~(\ref{eq:decay})
is shown in Figure~(\ref{fg:fourier}).  The amplitude of the various peaks, especially that of
$f_{\rm CBO} \equiv f_x$ has
a large amplitude relative to the white noise present in the spectrum,
implying that the CBO modulation is statistically very important.  The two 
CBO sidebands are
not of equal amplitude and in particular not equal to ${1 \over 2} A A_{\rm N}$, with
$ A_{\rm N}$ the amplitude at $f_x$.  This precludes that $N_0$ is the only CBO modulated
parameter as was assumed in the 1999 data analysis~\cite{brown}.

At early times
the number of positrons shown in Figure~(\ref{fg:wiggles})
is of order  10~million per 149.2~ns bin.  Therefore, very small beam dynamics
effects are important and noticeable in the least $\chi^2$ fits.  Since the acceptance 
of the 
detectors depends on the position of the muon beam relative to the detectors,
the time and energy spectra of the detected positrons are modulated with the
CBO frequency.  As a result, the g-2
phase, asymmetry and the normalization $N_0$ are all modulated with the CBO frequency, thus 
becoming all time dependent.  
Since the
CBO frequency is very close to  twice the g-2 frequency, it turns out~\cite{bennett} 
that the 
asymmetry and g-2 phase modulation are important effects that need special attention,
consistent with M.C. simulations.  The way they manifest themselves is by phase pulling
the g-2 frequency with an oscillation period of 
$T_o = {1 \over (f_x -f_a) - f_a} \approx 130 \, \rm \mu s$.  
The CBO modulation affects, as we said
earlier, the energy spectrum of the detected positrons and hence their average energy as a
function of time.  However, since the oscillation period is $\approx 130 \, \rm \mu s$, it
was difficult to distinguish it 
from other slow effects like gain change, muon losses, and pileup.  In 2001, we took data
at different $n$ values, specifically at $n=0.142$ and $n=0.122$ corresponding to a horizontal
CBO frequency of 491~kHz and 421~kHz, respectively.

The time dependence of the CBO modulated effects is given by:

\begin{enumerate}
\item{} $N_0(t) = N_0 \, [1 + A_N \, e^{-t/\tau_x} \, \cos{(2 \pi f_x t + \phi_N)}]$,
with $\tau_x$ the lifetime of the CBO modulation found from 
the data to be of the order of
$\approx 100 \rm \, \mu s$.

\item{} $A(t) = A_0 \, [1 + A_A \, e^{-t/\tau_x} \, \cos{(2 \pi f_x t + \phi_A)}]$,

\item{} $\phi_a(t) = \phi_{a0} + A_\phi \, e^{-t/\tau_x} \, \cos{(2 \pi f_x t + \phi_\phi)}$

\end{enumerate}

\begin{figure}[p]
\includegraphics*[width=18pc]{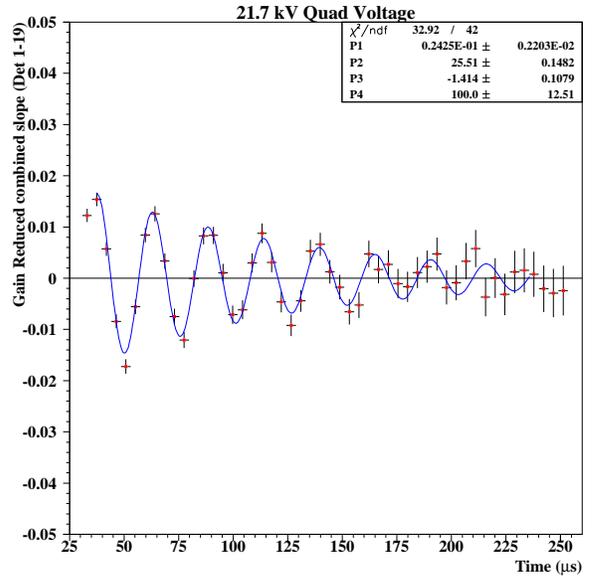}
\vskip -0.8 cm

\caption[beat1]{{By strobing the energy spectrum
of the detected electrons of the 2001 data we were able to observe the time dependence
of  $N_0(t)$.  The high voltage was 21.7~kV, and $n=0.122$.}
\label{fg:beat1}} 
\end{figure}

\begin{figure}[p]
\includegraphics*[width=18pc]{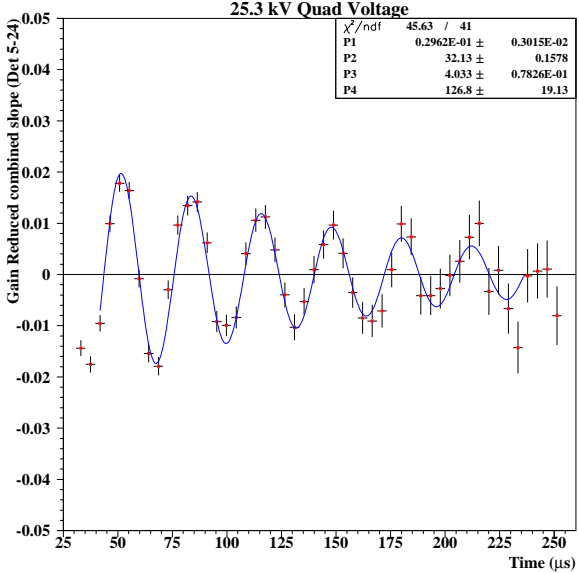}
\vskip -0.8 cm

\caption[beat2]{{Same as in Figure~(\ref{fg:beat1}) but for the case where 
the high voltage was 25.3~kV, and $n=0.142$.}
\label{fg:beat2}} 
\end{figure}

\noindent The time dependence of $N_0(t)$ was found by strobing  the
energy spectrum of the 2001 data at the g-2 frequency, see 
Figures~(\ref{fg:beat1},\ref{fg:beat2}).  The fitting function used was of the form:
$P_1 \, \sin{(2 \pi t / P_2 + P_3)} \, e^{-t/P_4}$.
Next it was verified by looking at the
residuals after the 5-parameter fit to the data, minimizing $\chi^2$ by fitting
the data with various functions, M.C. simulations, etc.  The time dependence used for
 $A(t)$ and $\phi(t)$ was not possible to verify from the data, only the M.C. simulations
showed that they could not be too far off.

The amplitudes of $A_N$, $A_A$, and $A_\phi$ are consistent with values from M.C. 
simulations.  The values of the phases $\phi_N$ versus detector from the fits to the data 
are consistent with running from 0 to $2 \pi$. That means that if the sum of all detectors
is used, the amplitudes of $A_N$, $A_A$, and $A_\phi$ are reduced substantially, consistent
with the values obtained with fits to the sum.

\begin{figure}[p]

\includegraphics*[width=17pc]{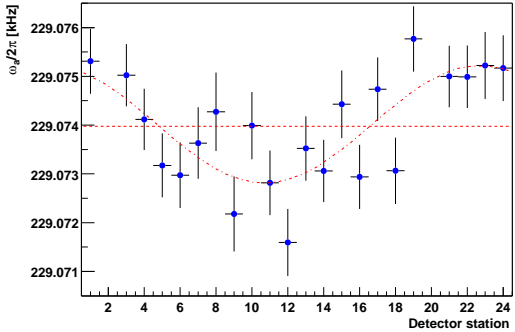}
\vskip -0.8 cm

\caption[rvsdet1]{{  A straight line fit to $\omega_a/2 \pi$ versus
detector number when only the ideal 5 parameter function is
used to fit the positron time spectrum.  
The $\chi^2$ for the straight line fit is $\chi^2 = 58.7 / 21$, and
the average $f_a = 229 \, 073.98 \pm 0.14$~Hz. A sine wave fit (see text) gives a good $\chi^2$ and a
central $f_a= 229 \, 074.02 \pm 0.14$~Hz, very close to the average.}
\label{fg:rvsdet1}} 
\end{figure}

The phase pulling of g-2 due to CBO is best depicted in Figure~(\ref{fg:rvsdet1}) where
a straight line fit to $f_a \equiv \omega_a/2 \pi$ 
versus detector gives $f_a = 229 \, 073.98 \pm 0.14$~Hz
and a $\chi^2 = 58.7 / 21$, indicating
that there is a consistency problem.  A fit to a sine wave plus a constant, 
$f_a + P_1 * \sin{({2 \pi \over 24} \times det.\# + P_2)}$, to the same data gives
$f_a = 229 \, 074.02 \pm 0.14$~Hz
and a $\chi^2 = 24.3 / 19$ with $P_1 =(1.20 \pm 0.20)$~Hz, and $P_2= 1.91 \pm 0.17$~rad.

\begin{figure}[p]
\includegraphics*[width=17pc]{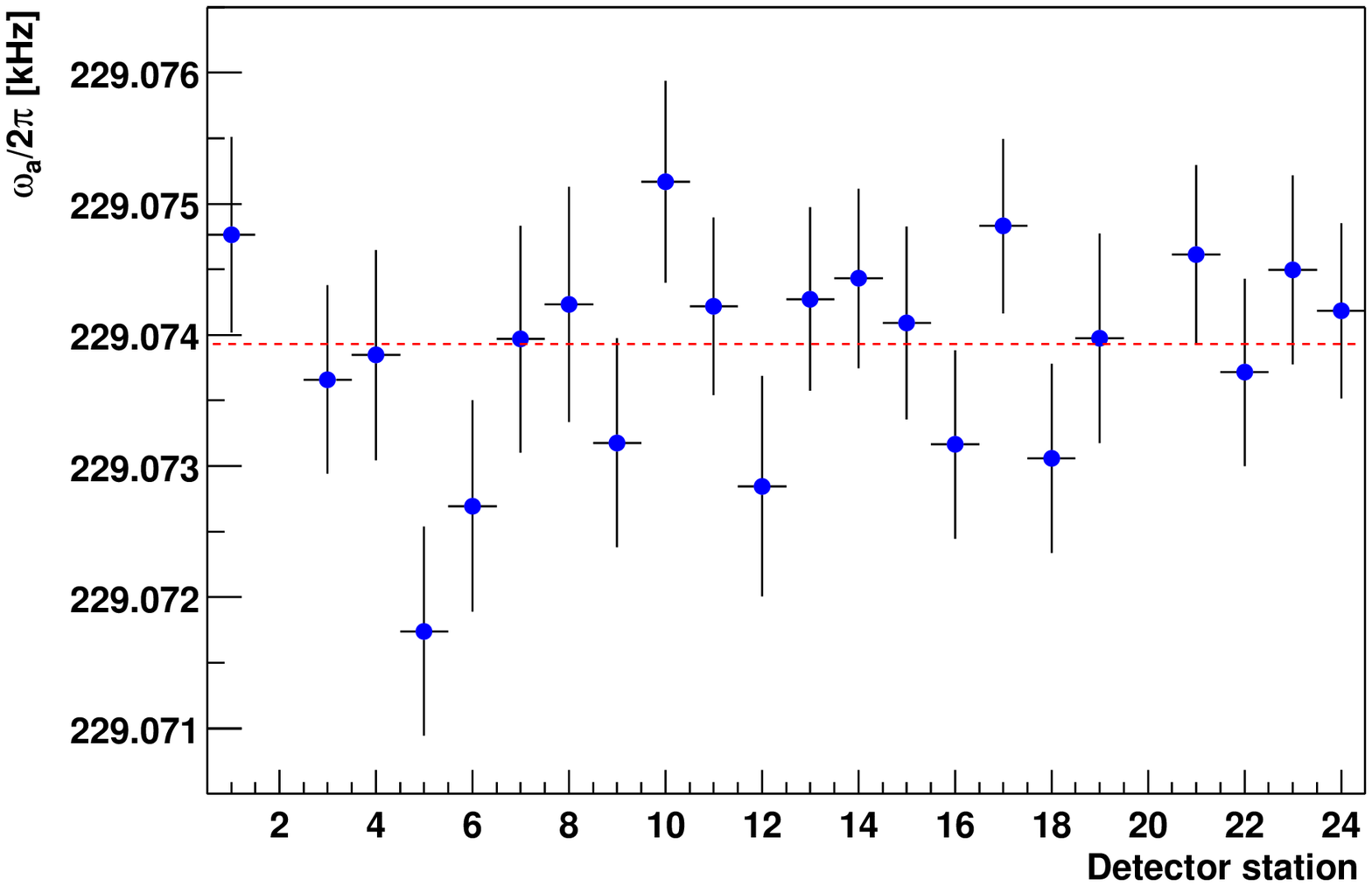}
\vskip -0.8 cm
\caption[rvsdet2]{{ A straight line fit to $\omega_a/2 \pi$ versus
detector number when the ideal 5 parameter function including the CBO modulations
of $N_0(t)$, $A(t)$, and $\phi(t)$, is
used to fit the positron time spectrum.  
The $\chi^2$ for the straight line fit is $\chi^2 = 23.9 / 21$, and
the average $f_a = 229 \, 073.92 \pm 0.14$~Hz.}
\label{fg:rvsdet2}} 
\end{figure}

When  the CBO modulation is included for the parameters
 $N_0(t)$, $A(t)$, and $\phi_a(t)$, 
the $\chi^2$ to a straight line fit
 of $f_a$ versus detector,  Figure~(\ref{fg:rvsdet2}),
is $\chi^2 = 23.9 / 21$, and
the average $f_a = 229 \, 073.92 \pm 0.14$~Hz.

\subsection{Different Approaches to the CBO Modulation and Slowly Varying Effects}
We have used several different approaches to analyze the data taken in 2000:

\begin{itemize}
\item{} Positrons with $E>2$~GeV and a 
function including the modulation of  $N_0(t)$ and $A(t)$ with $f_x$.

\item{} Positrons in 200~MeV energy bins with $1.4<E<3.2$~GeV and a 
function including the modulation of  $N_0(t)$, $A(t)$, and $\phi(t)$ with $f_x$.

\item{} Ratio method~\cite{brown}; $\omega_a$ becomes independent of slow effects, e.g. 
muon losses.

\end{itemize}

All the above methods gave results that are consistent among themselves
within the expected statistical
 uncertainties
due to the slightly different data used.  

In a side study, using positrons with $E>2$~GeV, we strobed 
the data at the horizontal CBO frequency $f_x$ making $\omega_a$ 
independent of the CBO parameters.
 Since $f_x$ is slightly higher than twice the g-2 frequency, one can recover all the 
information regarding the g-2 frequency, satisfying the Nyquist
limit. Therefore the frequency, amplitude, and phase are recovered using
mainly\footnote{There is also a slowly changing function multiplying the ideal function 
 describing slowly changing effects, like muon losses, detector gain change with time, etc. 
Those are included to improve the overall $\chi^2$ but make no difference in the final $f_a$
value obtained from the fits.}  only the 5-parameter function 
 and thus making no assumption as to the CBO functional form
whatsoever.  This method gave, again, consistent results with the above methods.

\subsection{Pileup and other Systematic Errors}

\begin{figure}[p]
\includegraphics*[width=18pc]{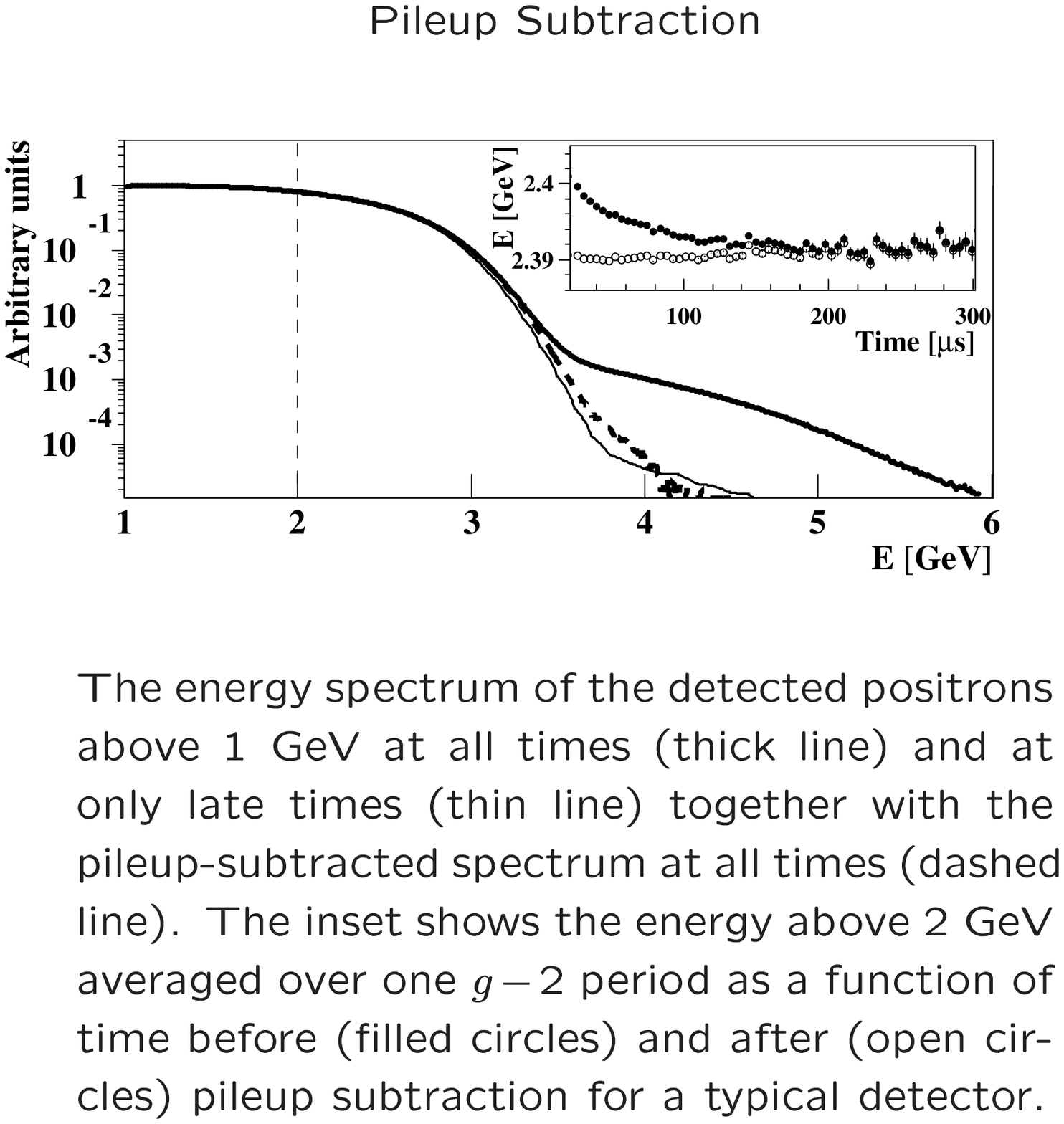}
\vskip -0.8 cm

\caption[pileup]{{The energy spectrum
of the detected positrons with energy greater that 1~GeV at all times (thick line)
and at only late times (thin line) when the rates are low.
The dashed line shows the pileup-subtracted spectrum at all times.  The inset 
shows the
average energy of positrons before (filled circles) and after (open circles)
pileup subtraction  for $E > 2$~GeV. }
\label{fg:pileup}} 
\end{figure}

When a positron arrives at the electromagnetic calorimeter and deposits 
an energy greater than  
approximately 1~GeV, it triggers the WFD connected to that particular detector.
The WFDs  are always running, and they were designed to keep 
in their memory
 more data, before and after the pulse,
than is necessary to reconstruct the positron signal that
triggered them.
This fact turned out to be of great help in dealing with the pileup pulses.
Due to high rates in 2000, the overlapping of two positron pulses
constitutes approximately $0.5\%$ of all detected pulses. We used the extra
recorded data to 
reconstruct, on a statistical basis, the time and energy spectrum of the
pileup pulses which we then subtract from the data; see 
Figure~(\ref{fg:pileup})~\cite{brown}.

The total systematic uncertainty in $\omega_a$ is 0.31~ppm~\cite{bennett}.
The final $f_a = 229 \, 074.11(14)(7)$~Hz (0.7~ppm)
which includes a total correction of +0.76(3)~ppm  due to 
electric field~\cite{picasso} and pitch~\cite{picasso,farley} corrections.
Those corrections have been studied in many different ways: analytically, particle
tracking and spin tracking.  For the latter, the BMT~\cite{telegdi} equations were applied
with our ring parameters, following the equations of chapter 11.11 (pages 556-560)
of reference~\cite{jackson}.  The results from all the above methods
are in agreement to  $\approx 0.01$~ppm.


\section{Analysis of $\omega_p$}

\begin{figure}[p]
\includegraphics*[width=18pc]{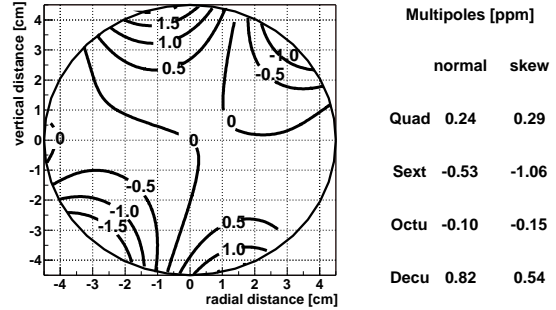}
\vskip -0.8 cm
\caption[field]{{ The  magnetic field multipoles 
integrated over the  azimuth of the ring for one trolley run.}
\label{fg:field}} 
\end{figure}

The magnetic field of the muon storage region
 was measured with an NMR trolley every two to three days while in between it was
followed by 367 fixed NMR probes located on top and bottom of the vacuum chambers.
The average B-field, convoluted over the muon distribution, was obtained by
 two largely independent analyses of   $\omega_p$~\cite{bennett}.  The magnetic
field multipoles integrated over the  azimuth of the ring for one trolley run
out of 22 are shown in Figure~(\ref{fg:field}); 
the central field was $1.451\, 274$~T.   
The final $\omega_p/(2 \pi) = 61 \, 791 \, 595 (15)$~Hz. 
The total systematic uncertainty in $\omega_p$ is 0.24~ppm~\cite{bennett}.

\section{Results}

In order to compute $a_\mu$, both $\omega_a$ and $\omega_p$ values are 
necessary.
The analysis groups dealing with $\omega_a$ and $\omega_p$ worked separately
and had applied secret offsets to their results until it was decided the 
analyses were finished.  This avoided  biases that could influence the
choice of data selection, analysis method, etc.  After the analyses seemed complete,
during a collaboration meeting  a secret ballot was held of 
whether or not we should reveal the offsets and compute $a_\mu$.  It was
unanimous for revealing the offsets and compute  $a_\mu$, which we did.
The results are 

\begin{equation}
a_{\mu^+} = {R \over \lambda - R} = 11 \, 659 \, 204 (7)(5) \times 10^{-10}, \label{eq:res1}
\end{equation}

\noindent with a relative error of 0.7~ppm.  The experimental world average becomes

\begin{equation}
a_\mu(exp) = 11 \, 659 \, 203 (8) \times 10^{-10}, \label{eq:res2}
\end{equation}

\noindent with again a relative error of 0.7~ppm.  This experiment, like
most of the muon experiments, is still statistics
limited.

\begin{figure}[p]
\includegraphics*[width=19.5pc]{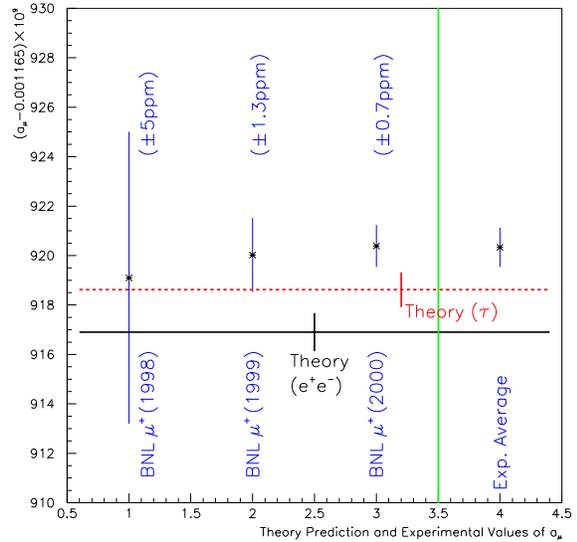}
\vskip -0.8 cm
\caption[results3]{{ The experimental values of  $a_\mu$ compared to the
theoretical  $a_\mu$ based on the current standard model  
with the $e^+ e^-$ data (solid horizontal line) and
the $\tau$ data (dashed line)  for the hadronic contribution~\cite{davier2}.}
\label{fg:results3}} 
\end{figure}

 In Figure~(\ref{fg:results3}) we give the recent BNL $a_\mu(exp)$ values, the 
average and the theoretical values based on the current standard model 
with the $e^+ e^-$ data (solid horizontal line) and
the $\tau$ data (dashed line)  for the hadronic contribution~\cite{davier2}.

\section{Discussion and Future Prospects}
The difference between the  experimental value and  the current theoretical prediction of 
 $a_\mu$ is 

\begin{equation}
 a_\mu({\rm exp})- a_\mu({\rm SM}) = 34 (11) \times 10^{-10}
\end{equation}

\noindent which is  a little over
three standard deviations and it may indicate new physics. One should, however,
 wait for confirmation of the $e^+e^-$ data and understand the reason why
the $\tau$ based data
imply a higher hadronic contribution than the $e^+e^-$ data,
before making any claims as to whether or not
new physics has been seen.   The difference 
$ a_\mu({\rm exp})- a_\mu({\rm SM}) = 17 (11) \times 10^{-10}$, i.e. only 1.5 sigma,
 when the $\tau$ data are used.
 On the experimental side, we have already accumulated
about 3~billion electrons with $E > 2$~GeV, equivalent to a statistical power of 
approximately 0.7~ppm, from our 2001 run with negative muons.
 We are currently 
analyzing those data and expect to finish by early next year.  We also have 
 scientific approval for an extra four month period which the High Energy Division
of DOE has not yet approved, though 
they should, in order to properly conclude the experiment.

Assuming that the $e^+e^-$ data  will hold, and
if supersymmetry is responsible for 
the g-2 deviation, an EDM from similar quantum loops but with a phase giving 
rise to T and P violation is natural and likely.  That would make the muon one 
of the best places to search for an EDM~\cite{marciano2}.  Such an effort 
is currently underway~\cite{edm}. 
It promises to be sensitive to physics beyond the standard 
model~\cite{pilaftsis,babu,matchev,ellis,strumia} and to continue the exciting muon
physics of the past and present.

\end{document}